# A peculiarity of localized mode transfiguration of a Cantor-like chiral multilayer

**Vladimir R. Tuz**

Department of Theoretical Radio Physics, Kharkov National University, Svobody Square 4, Ukraine

E-mail: tvr@vega.kharkov.ua, Vladimir.R.Tuz@univer.kharkov.ua

**Abstract**

A fractal-like (Cantor-like) stratified structure of chiral and convenient isotropic layers is considered. Peculiarities of the wave localization, self-similarity, scalability and sequential splitting in the reflected field of both the co-polarized and cross-polarized components are studied. The appearing of the additional peak multiplets in stopbands is revealed, and a correlation of their properties with chirality parameter is established.

(Some figures in this article are in color only in the electronic version)

**1. Introduction**

Inhomogeneous regular media whose optical properties are varied in space in a predetermined way are subjects of study of numerous research papers. A broad class of such regular systems is stratified structures whose properties vary periodically along a certain preferential direction. It is conventional in optics to refer such structures to photonic crystals [1, 2]. The theory of propagation of electromagnetic waves in the photonic crystals is constructed in analogy with the theory of motion of electrons in a periodic potential in solids. On this basis, the notions of eigenwaves of the wave vectors of plane monochromatic waves, the forbidden zones in the spectrum of eigenvalues, the normal waves, etc. were introduced. However, this analogy occasionally is inadequate especially in case of optically active media which are intensively applied to the operation in photonics in last years. The sequential solution of the diffraction problem is required in this case.

One class of the optically active media is chiral materials. They are characterized by an intrinsic left- or right-handedness at optical frequencies, due to a helical structure. Renewed, recent interest in chiral media stems primarily from the possibility that, through advances in polymer science, or through the manufacturing of artificial dielectrics, substances possessing a rotatory power in optics might be produced [3-10]. When chiral materials applied to the photonic crystals, the latter hold rich optical properties and are characterized by the selective reflection in a specific wavelength region in addition to circular polarization of the reflected and transmitted light. These properties are attractive in a design of cascaded high-$Q$ and stopband frequency filters, polarizers, high-precision matched loads, etc. Specifically the using of chiral materials in distributed feedback lasers can increase the distributed feedback laser coupling strength and reduce the laser threshold gain for certain ranges of material properties [11].

On the other hand, particular emphasis in a photonic crystal family has been placed on structures with periodicity defects [1-2]. It is well known that the defect inside a stratified isotropic one-dimensional sample produces additional defective modes (localized modes) within stopbands. These localized modes are employed in the construction of distributed feedback lasers with very narrow bandwidths. Generally a configuration of the localized modes is depended on the material parameters of the defective element and its position within the structure [12]. It is apparent that the introduction of plural defects inside a photonic crystal can significantly change the localized mode features. Optical properties of deterministic aperiodical (quasi-periodical) structures are a good case in point [13-20]. It is conventional to separate these structures into two major groups accordingly to their construction algorithm, namely, substitutional and model fractal multilayers [18]. In either case the strong wave localization, the scalability and self-similarity of the transmitted and reflected spectra are revealed.

In analogy with isotropic periodic structures, a defect in a chiral sample can be produced by inserting an isotropic or anisotropic layer into the sequence. In addition, a "chiral twist" defect can be created by rotating helical axis of the portion of existing layers [21-25] or by inserting layers with other optical activity strength and handedness [26, 27]. At the present time the theory of multilayered chiral photonic structures with plural defects or quasi-periodically organized is not sufficiently advanced, although, by virtue of the unique behaviors of chiral materials, their optical properties will possess some peculiarities in contrast to achiral ones [28]. They can expand the field of applications of stratified chiral structures, as example, it has been recently reported in [29] that the introducing the Fibonaccian defects into the chiral photonic crystals can create reflective color-displays without the need for back-lighting, polarizers, or color filters.

In this paper we should focus an attention on some behaviors of the localized mode configuration in a quasi-periodical fractal-like structure with chiral layers. The diffraction problem solution is obtained using the $2 \times 2$-block representation transfer matrix formulation [10, 26-28].

## 2. Problem formulation

Let us to consider a fractal-like stratified structure that is generated in a way similar to the triadic Cantor set construction [18, 30]. The Cantor multilayer is characterized by two fundamental parameters, the generator $G = 3, 5, 7, ...$ and the generation number $N = 1, 2, 3, ...$ . Sample structures are shown in Fig. 1(a), and the stack construction method can be understood from there. At first stage growth, starting with an interval $[0, \Lambda]$ ($\Lambda = GL$ is the total thickness of the structure) the certain parts are removed, forming a Cantor set of order $N = 1$ composed of the subsets $[0, L], ..., [(G-1)L, \Lambda]$ (green gaps) that separated by intervals $[L, 2L], ... [(G-2)L, (G-1)L]$ (yellow gaps). The Cantor set of order $N = 2$ is obtained by removing again the certain parts of these subsets resulting in the subsets $\{[0, L/G], [2L/G, 3L/G], ..., [(G-1)L/G, L]\}$, $\{[2L, (2G+1)L/G], [(2G+2)L/G, (2G+3)L/G], ..., [(3G-1)L/G, 3L]\}$, ..., $\{[(G-1)L, ((G-1)+1)L/G], [((G-1)+2)L/G, ((G-1)+3)L/G], ..., [(G-1)(G+1)L/G, \Lambda]\}$. High-order sets are formed in similar ways. It can be observed that the subsets are copies of the original set scaled by the factor of $1/G$. This property is called self-similarity. Two other properties are the recursive property and the formation of fine structures [16]. The recursive property comes form the fact that the certain parts of each subset is removed when the order increases from $N$ to $N+1$. The fine structure comes from the specific rule of the subset formation with the result that it is possible to know the form of each subsets at any orders $N$. The fractal dimensionality of a Cantor set of any order $N$ attributed to the structure under study is $\ln\left[(G+1)/2\right]/\ln G$ [18].

The layers making up the structure are homogeneous, isotropic, and infinite in two transverse directions. The yellow layers are magnetodielectric (material) slabs with permittivity $\varepsilon_1$ and permeability $\mu_1$, and the green ones are chiral slabs with $\varepsilon_2$, $\mu_2$ and chirality parameter $\rho$, respectively [Fig. 1(b)]. In general, the material parameters are frequency dependent and complex for lossy media $\varepsilon_j(\omega) = \varepsilon_j'(\omega) + i\varepsilon_j''(\omega)$, $\mu_j(\omega) = \mu_j'(\omega) + i\mu_j''(\omega)$, ($j = 1, 2$), $\rho(\omega) = \rho'(\omega) + i\rho''(\omega)$. Here $\rho'$ is responsible for optical rotatory power and $\rho''$ produces circular dichroism. The outer half-spaces ($z \leq 0$, $z \geq \Lambda$) are homogeneous, isotropic with material parameters $\varepsilon_0$ and $\mu_0$.

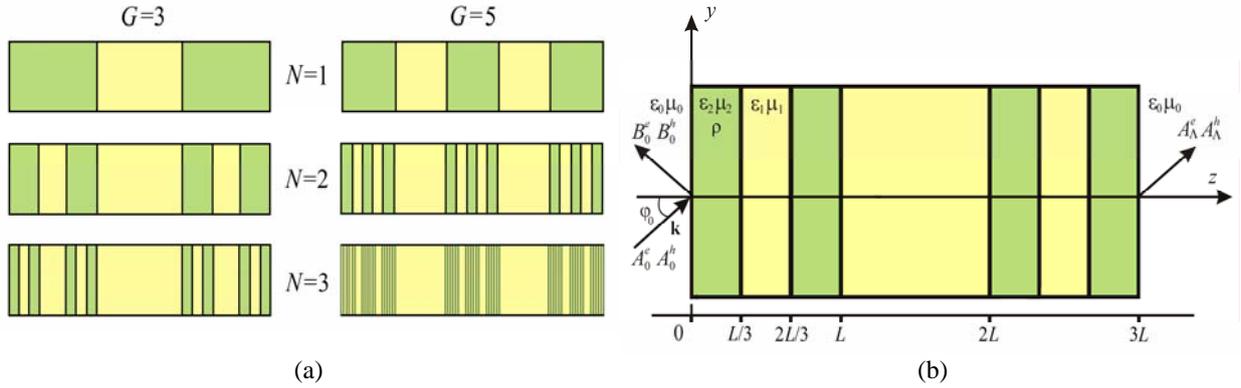

(a)                           (b)

**Figure 1. (color online)** (a) Illustration of the generation of a self-similar Cantor set at different stages $G$ and $N$. (b) Geometry of reflection and transmission problem of a Cantor-like ($G = 3$, $N = 2$) structure of chiral and material layers.

As the excitation fields, the plane (in the $YOZ$ plane) monochromatic waves with perpendicular ($\vec{E}^e \parallel \vec{x}_0$, $H_x^e = 0$) or parallel ($\vec{H}^h \parallel \vec{x}_0$, $E_x^h = 0$) polarization are selected ($E$– and $H$–waves). They are obliquely incident from the region $z \leq 0$ at an angle $\varphi_0$ to the $z$-axis.

$$\begin{Bmatrix} E_{x0}^e \\ E_{y0}^h \end{Bmatrix} = \pm \begin{Bmatrix} A_0^e / \sqrt{Y_0^e} \\ iA_0^h / \sqrt{Y_0^h} \end{Bmatrix} \exp[-i(\omega t - k_{y0} y - k_{z0} z)], \quad \begin{Bmatrix} H_{y0}^e \\ H_{x0}^h \end{Bmatrix} = \begin{Bmatrix} A_0^e \sqrt{Y_0^e} \\ iA_0^h \sqrt{Y_0^h} \end{Bmatrix} \exp[-i(\omega t - k_{y0} y - k_{z0} z)], \quad (1)$$

where $k_{z0} = k_0 \cos\varphi_0$, $k_{y0} = k_0 \sin\varphi_0$, $k_0 = n_0 k$, $k = \omega/c$ is the free-space wavenumber, $n_0 = \sqrt{\varepsilon_0 \mu_0}$, and $Y_0^e = Z_0^{-1} \cos\varphi_0$, $Y_0^h = (Z_0 \cos\varphi_0)^{-1}$ are the wave admittances of the input and output half-spaces, and $Z_0 = \sqrt{\mu_0/\varepsilon_0}$.

### 3. Fields in a chiral layer

In the homogeneous along the $x$ direction chiral layer, the electromagnetic field is characterized by the general displacement [31]

$$\vec{D} = \varepsilon_2 \vec{E} + i\rho \vec{H}, \quad \vec{B} = \mu_2 \vec{H} - i\rho \vec{E}, \quad (2)$$

and Maxwell's equations written as

$$\text{curl}\vec{E} = ik\vec{B}, \quad \text{curl}\vec{H} = -ik\vec{D} \quad (3)$$

from which it follows that $\text{div}\vec{B} = 0$ and $\text{div}\vec{D} = 0$.

Taking the curl operator on both sides of (3) and using the constitutive equations (2) yields the next two coupled differential equations related to the field components $E_x$ and $H_x$

$$\Delta_\perp E_x + k^2(n_2^2 + \rho^2) E_x - 2ik^2 \rho \mu_2 H_x = 0, \quad \Delta_\perp H_x + k^2(n_2^2 + \rho^2) H_x + 2ik^2 \rho \varepsilon_2 E_x = 0, \quad (4)$$

in which $n_2 = \sqrt{\varepsilon_2 \mu_2}$ is the refractive index of the chiral medium, and $\Delta_\perp = \partial^2/\partial y^2 + \partial^2/\partial z^2$ is the two-dimensional Laplacian.

It is well known that the normal waves of an unbounded chiral medium are right ($\vec{Q}^{s+}$) and left ($\vec{Q}^{s-}$) circularly polarized eigenwaves, thus the field components of the waves of the perpendicular and parallel linear polarizations can be presented as the superposition of these two circularly polarized eigenwaves [31]:

$$E_{x2}^e = Q^{e+} + Q^{e-}, \quad H_{x2}^e = -iZ_2^{-1}(Q^{e+} - Q^{e-}),$$
$$E_{x2}^h = iZ_2(Q^{h+} - Q^{h-}), \quad H_{x2}^h = Q^{h+} + Q^{h-}, \quad (5)$$

where $Z_2 = \sqrt{\mu_2/\varepsilon_2}$ is the wave impedance of the chiral medium. Such substitution transforms Eq. (4) into two independent Helmholtz equations:

$$\Delta_\perp Q^{s\pm} + \left(\gamma^\pm\right)^2 Q^{s\pm} = 0. \tag{6}$$

Here $s = e, h$; $\gamma^\pm = k\sqrt{\varepsilon_2^\pm \mu_2^\pm} = k n_2^\pm = k(n_2 \pm \rho)$ are the propagation constants of the right ($\gamma^+$) (RCP) and left ($\gamma^-$) (LCP) circularly polarized eigenwaves, respectively, in the unbounded chiral media with the equivalent material parameters $\varepsilon_2^\pm = \varepsilon_2 \pm \rho Z_2^{-1}$ and $\mu_2^\pm = \mu_2 \pm \rho Z_2$ [31]. The general solutions of (6) for the RCP and LCP waves in a bounded chiral layer can be expressed by

$$\begin{Bmatrix} Q^{e\pm} \\ Q^{h\pm} \end{Bmatrix} = \begin{Bmatrix} 1/2\sqrt{Y_2^{e\pm}} \\ \sqrt{Y_2^{h\pm}}/2 \end{Bmatrix} \left[ \begin{Bmatrix} A_2^{e\pm} \\ A_2^{h\pm} \end{Bmatrix} \exp\left[i\left(\gamma_{y2}^\pm y + \gamma_{z2}^\pm z\right)\right] + \begin{Bmatrix} B_2^{e\pm} \\ B_2^{h\pm} \end{Bmatrix} \exp\left[i\left(\gamma_{y2}^\pm y - \gamma_{z2}^\pm z\right)\right] \right], \tag{7}$$

where $A_2^{s\pm}$, $B_2^{s\pm}$ denote the field amplitudes, $Y_2^{e\pm} = Z_2^{-1}\cos\varphi_2^\pm$, $Y_2^{h\pm} = \left(Z_2 \cos\varphi_2^\pm\right)^{-1}$ are the wave admittances, $\gamma_{y2}^\pm = \gamma^\pm \sin\varphi_2^\pm$, $\gamma_{z2}^\pm = \gamma^\pm \cos\varphi_2^\pm$, and $\varphi_2^\pm = \sin^{-1}\left[n_0 \sin\varphi_0 / n_2^\pm\right]$ are the refracted angles of the two eigenwaves in the chiral layers. The substitution of Eq. (7) into Eq. (5) gives the field components of the $e$ and $h$ polarizations.

## 3. Transfer matrix. Reflected and transmitted fields

When a system includes chiral layers, the four magnetic and electric parts of the plane wave are spatially dependent to each other, and a so-called mode coupling appears. Making use of the transfer-matrix formalism [32], the equations coupling the field amplitudes at the structure input $\mathbf{V}_0 = \begin{pmatrix} A_0^h & B_0^h & A_0^e & B_0^e \end{pmatrix}^T$ and output $\mathbf{V}_\Lambda = \begin{pmatrix} A_\Lambda^h & 0 & A_\Lambda^e & 0 \end{pmatrix}^T$ ($T$ is the matrix transpose operator) are written as [10, 26-28]

$$\mathbf{V}_0 = \mathbf{T}_\Sigma \mathbf{V}_\Lambda = \left\{ \mathbf{T}_g(d)\mathbf{T}_y(d)\mathbf{T}_g(d)...\mathbf{T}_g(d) \right\} \mathbf{V}_\Lambda, \tag{8}$$

where the total transfer matrix $\mathbf{T}_\Sigma$ is obtained by multiplying in the appropriate order the matrices corresponding to each layer in the structure.

The matrices $\mathbf{T}_y$ and $\mathbf{T}_g$ are the particular transfer matrices of rank 4 of the material (yellow) and chiral (green) layers of the certain thickness $d$, respectively. They are

$$\mathbf{T}_y(d) = \mathbf{T}_{01}\mathbf{P}_1(d)\mathbf{T}_{10} = \begin{pmatrix} \left(\mathbf{T}_y^{hh}\right) & 0 \\ 0 & \left(\mathbf{T}_y^{ee}\right) \end{pmatrix}, \quad \mathbf{T}_g(d) = \mathbf{T}_{02}\mathbf{P}_2(d)\mathbf{T}_{20} = \begin{pmatrix} \left(\mathbf{T}_g^{hh}\right) & \left(\mathbf{T}_g^{he}\right) \\ \left(\mathbf{T}_g^{eh}\right) & \left(\mathbf{T}_g^{ee}\right) \end{pmatrix}, \tag{9}$$

where $\mathbf{T}_{0j}$ and $\mathbf{T}_{j0}$ ($j = 1,2$) are the transfer-matrices of the layer interfaces with outer half-spaces, and $\mathbf{P}_j(d)$ are the propagation matrices through the corresponding layer. The elements of the block-matrices $\mathbf{T}_{0j}$ and $\mathbf{T}_{j0}$ are determined by solving the boundary-value problem related to the field components (1) and (5). In the block representation ($2 \times 2$) [28] the transfer-matrices of the material (yellow) layer are

$$\mathbf{T}_{01} = \begin{pmatrix} \left(\mathbf{T}_{01}^{hh}\right) & 0 \\ 0 & \left(\mathbf{T}_{01}^{ee}\right) \end{pmatrix}, \quad \mathbf{T}_{10} = \begin{pmatrix} \left(\mathbf{T}_{10}^{hh}\right) & 0 \\ 0 & \left(\mathbf{T}_{10}^{ee}\right) \end{pmatrix}, \quad \mathbf{P}_1 = \begin{pmatrix} \left(\mathbf{E}_1\right) & 0 \\ 0 & \left(\mathbf{E}_1\right) \end{pmatrix} \tag{10}$$

$$\mathbf{T}_{pv}^{ss} = \frac{1}{2\sqrt{Y_p^s Y_v^s}} \begin{pmatrix} Y_p^s + Y_v^s & \pm(Y_p^s - Y_v^s) \\ \pm(Y_p^s - Y_v^s) & Y_p^s + Y_v^s \end{pmatrix}, \quad \mathbf{E}_1(d) = \begin{pmatrix} \exp(-ik_{z1}d) & 0 \\ 0 & \exp(ik_{z1}d) \end{pmatrix}, \tag{11}$$

where $\mathbf{T}^{ss}_{pv}$ corresponds to the matrices $\mathbf{T}^{hh}_{01}$, $\mathbf{T}^{ee}_{01}$ and $\mathbf{T}^{hh}_{10}$, $\mathbf{T}^{ee}_{10}$, and the upper sign relates to $s = h$, and the lower sign relates to $s = e$ in terms of the wave types, $k_{z1} = kn_1 \cos\varphi_1$, $n_1 = \sqrt{\varepsilon_1 \mu_1}$, $\varphi_1 = \sin^{-1}[n_0 \sin\varphi_0 / n_1]$.

The transfer-matrices of the chiral (green) layer are

$$\mathbf{T}_{02} = \begin{pmatrix} (\mathbf{T}^{hh}_{02+}) & (\mathbf{T}^{hh}_{02-}) \\ (\mathbf{T}^{ee}_{02+}) & (\mathbf{T}^{ee}_{02-}) \end{pmatrix}, \quad \mathbf{T}_{20} = \begin{pmatrix} (\mathbf{T}^{hh}_{20+}) & (\mathbf{T}^{ee}_{20+}) \\ (\mathbf{T}^{hh}_{20-}) & (\mathbf{T}^{ee}_{20-}) \end{pmatrix}, \quad \mathbf{P}_2 = \begin{pmatrix} (\mathbf{E}^+_2) & 0 \\ 0 & (\mathbf{E}^-_2) \end{pmatrix} \quad (12)$$

$$\mathbf{T}^{hh}_{02\pm} = \frac{1}{4\sqrt{Y^h_0 Y^{h\pm}_2}} \begin{pmatrix} Y^{h\pm}_2 + Y^h_0 & Y^{h\pm}_2 - Y^h_0 \\ Y^{h\pm}_2 - Y^h_0 & Y^{h\pm}_2 + Y^h_0 \end{pmatrix}, \quad \mathbf{T}^{ee}_{02\pm} = \mp \frac{1}{4\sqrt{Y^e_0 Y^{e\pm}_2}} \begin{pmatrix} Y^e_0 + Y^{e\pm}_2 & Y^e_0 - Y^{e\pm}_2 \\ Y^e_0 - Y^{e\pm}_2 & Y^e_0 + Y^{e\pm}_2 \end{pmatrix},$$

$$\mathbf{T}^{hh}_{20\pm} = \frac{1}{4Y^{h\mp}_2 \sqrt{Y^h_0 Y^{h\pm}_2}} \times$$

$$\begin{pmatrix} (Y^{h\mp}_2 + Y^h_0)(Y^{h\mp}_2 + Y^{h\pm}_2) - (Y^{h\mp}_2 - Y^h_0)(Y^{h\mp}_2 - Y^{h\pm}_2) & (Y^{h\mp}_2 + Y^h_0)(Y^{h\mp}_2 - Y^{h\pm}_2) - (Y^{h\mp}_2 - Y^h_0)(Y^{h\mp}_2 + Y^{h\pm}_2) \\ (Y^{h\mp}_2 + Y^h_0)(Y^{h\mp}_2 - Y^{h\pm}_2) - (Y^{h\mp}_2 - Y^h_0)(Y^{h\mp}_2 + Y^{h\pm}_2) & (Y^{h\mp}_2 + Y^h_0)(Y^{h\mp}_2 + Y^{h\pm}_2) - (Y^{h\mp}_2 - Y^h_0)(Y^{h\mp}_2 - Y^{h\pm}_2) \end{pmatrix}, \quad (13)$$

$$\mathbf{T}^{ee}_{20\pm} = \mp \frac{1}{4Y^{e\mp}_2 \sqrt{Y^e_0 Y^{e\pm}_2}} \times$$

$$\begin{pmatrix} (Y^{e\mp}_2 + Y^e_0)(Y^{e\mp}_2 + Y^{e\pm}_2) - (Y^{e\mp}_2 - Y^e_0)(Y^{e\mp}_2 - Y^{e\pm}_2) & (Y^{e\mp}_2 - Y^e_0)(Y^{e\mp}_2 + Y^{e\pm}_2) - (Y^{e\mp}_2 + Y^e_0)(Y^{e\mp}_2 - Y^{e\pm}_2) \\ (Y^{e\mp}_2 - Y^e_0)(Y^{e\mp}_2 + Y^{e\pm}_2) - (Y^{e\mp}_2 + Y^e_0)(Y^{e\mp}_2 - Y^{e\pm}_2) & (Y^{e\mp}_2 + Y^e_0)(Y^{e\mp}_2 + Y^{e\pm}_2) - (Y^{e\mp}_2 - Y^e_0)(Y^{e\mp}_2 - Y^{e\pm}_2) \end{pmatrix}.$$

The blocks of the quasi-diagonal propagation-matrix $\mathbf{P}_2$ are

$$\mathbf{E}^\pm_2(d) = \begin{pmatrix} \exp(-i\gamma^\pm_{z2} d) & 0 \\ 0 & \exp(i\gamma^\pm_{z2} d) \end{pmatrix}. \quad (14)$$

The behaviors and functional capabilities of the structure appear as properties of the reflection and transmission coefficients. For their determination, the equations coupling the field amplitudes at the structure input and output for the incident fields of the $h$-type ($A^e_0 = 0$) and the $e$-type ($A^h_0 = 0$) are written as:

$$\begin{pmatrix} A^h_0 & B^h_0 & 0 & B^e_0 \end{pmatrix}^T = \mathbf{T}_\Sigma \begin{pmatrix} A^h_\Lambda & 0 & A^e_\Lambda & 0 \end{pmatrix}^T, \quad \begin{pmatrix} 0 & B^h_0 & A^e_0 & B^e_0 \end{pmatrix}^T = \mathbf{T}_\Sigma \begin{pmatrix} A^h_\Lambda & 0 & A^e_\Lambda & 0 \end{pmatrix}^T. \quad (15)$$

The reflection and transmission coefficients of the reflected ($z \leq 0$) and transmitted ($z \geq \Lambda$) fields are determined by the expressions $R^{ss} = B^s_0 / A^s_0$, $R^{ss'} = B^{s'}_0 / A^s_0$ and $\tau^{ss} = A^s_\Lambda / A^s_0$, $\tau^{ss'} = A^{s'}_\Lambda / A^s_0$, respectively. From (15) they are:

$$\begin{aligned} R^{hh} &= (t_{21}t_{33} - t_{23}t_{31})/\Delta, & \tau^{hh} &= t_{33}/\Delta, \\ R^{he} &= (t_{41}t_{33} - t_{43}t_{31})/\Delta, & \tau^{he} &= -t_{31}/\Delta, \\ R^{ee} &= (t_{11}t_{43} - t_{41}t_{13})/\Delta, & \tau^{ee} &= t_{11}/\Delta, \\ R^{eh} &= (t_{11}t_{23} - t_{21}t_{13})/\Delta, & \tau^{eh} &= -t_{13}/\Delta, \end{aligned} \quad (16)$$

where $\Delta = t_{11}t_{33} - t_{31}t_{13}$, and $t_{pq}$ are the elements of the transfer matrix $\mathbf{T}_\Sigma$.

The direct transfer matrix product in (8) is computationally demanding, therefore, in view of the fact of the self-similarity of the fractal structure, the calculation algorithm for the determination of the total transfer matrix $\mathbf{T}_\Sigma = \mathbb{T}_N$ is constructed iteratively by the next formulas

$$\begin{cases} \mathbb{T}_1 = \left[ \mathbf{T}_g\left(L/G^{(N-1)}\right)\mathbf{T}_y\left(L/G^{(N-1)}\right)\right]^{(G-1)/2} \mathbf{T}_g\left(L/G^{(N-1)}\right), \\ \mathbb{T}_n = \left[ \mathbb{T}_{n-1}\mathbf{T}_y\left(L/G^{(N-n)}\right)\right]^{(G-1)/2} \mathbb{T}_{n-1}, \end{cases} \quad (17)$$

where $n = 2\ldots N$ is the iteration number. In the special situation $N = 0$, the total transfer matrix is $\mathbf{T}_\Sigma = \mathbb{T}_0 = \mathbf{T}_y(\Lambda)$, $\Lambda = GL$.

## 4. Numerical results. Solution analysis

It is well known [31] that the normal waves of an unbounded chiral medium are left- and right-hand circularly polarized eigenwaves which differ by the propagation constants $\gamma^\pm = k\left(n_2 \pm \rho\right)$, and each of this eigenwaves sees the medium as if it were an isotropic medium with different refraction indexes $n^\pm = n_2 \pm \rho$. A linearly polarized plane wave propagating through the chiral medium experiences a rotation of its polarization ellipse. As result, the cross-polarized component appears in the transmitted field at the output of the finite chiral structure (Fig. 2). The theoretical angle of the polarization rotation in a chiral medium is given by $\alpha = -\operatorname{Re}(\rho)kD$ [31], where $\operatorname{Re}(\rho)$ is the real part of the chirality parameter, $k$ is the wavenumber in free space, and $D = (G+1)^N L/(2^N G^{N-1})$ is the distance passed by the wave through chiral layers.

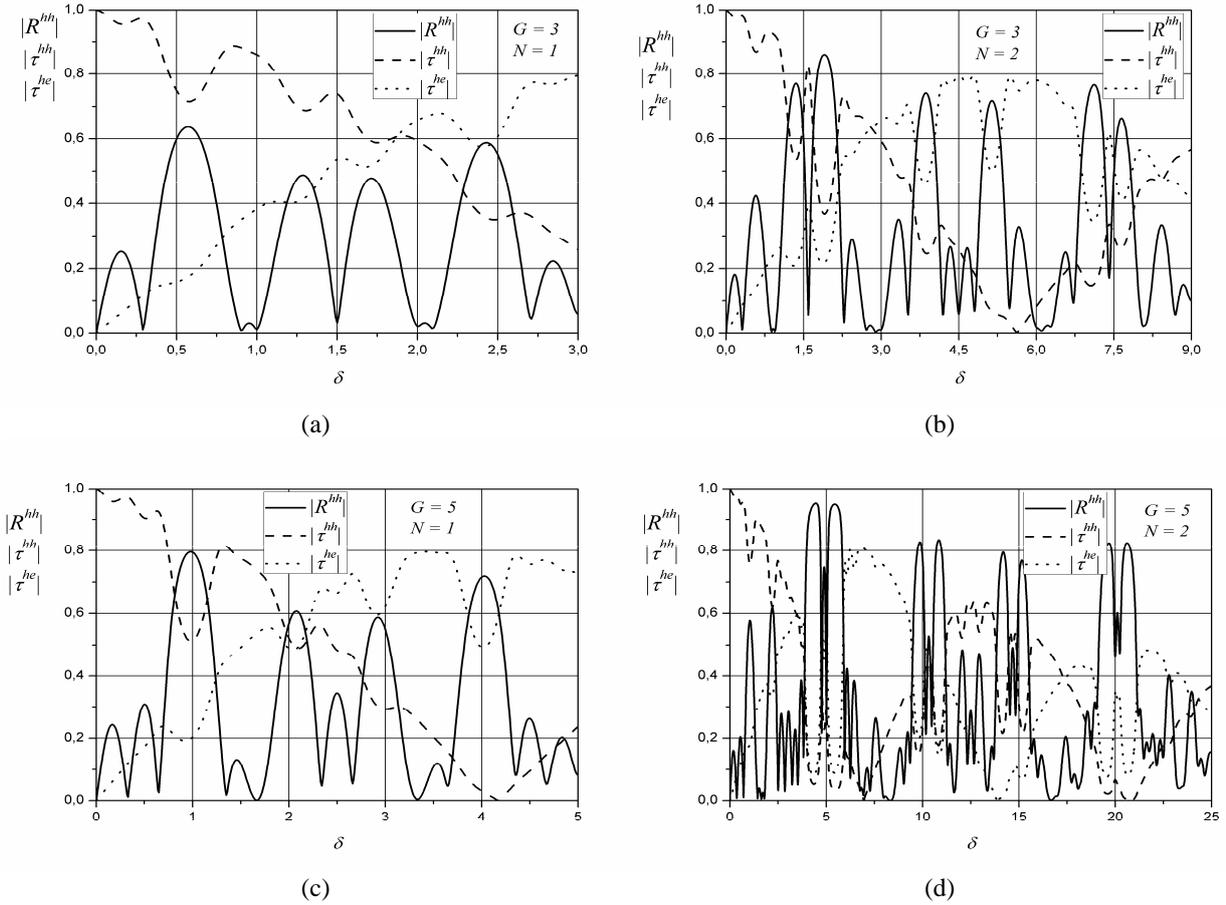

**Figure 2.** The amplitude of the reflection and transmission coefficients of a Cantor-like multilayer with chiral layers as a function of the thickness-to-wavelength ratio $\delta$ for different stages of fractal growth $(G, N)$.

$n_0 = n_1 = 1$, $n_2 = 1.5 + 0.013i$, $\rho = 0.1$, $\varphi_0 = 0^0$.

The amplitude of the reflection and transmission coefficients (the reflection and transmission spectra) is a periodic function (with the period divisible by $G$) of the thickness-to-wavelength ratio $\delta = L/\lambda$ with alternating bands of high and low average level of the reflection (the stopbands and the passbands). In a presence of small real losses ($\varepsilon_j'' \neq 0$, $\mu_j'' \neq 0$, $j = 1,2$) of layers, the average level of the reflection and transmission reduces, and the amplitude of small-scale oscillations in the passbands decreases. The spectra have self-similar properties i.e. reflection (transmission) coefficient variation at each higher stage is modulated version of that associated with the previous stage [15]. In a case of the normal incidence ($\varphi_0 = 0^0$) a linearly polarized wave that travels forward trough a chiral layer with some thickness, and travels back after reflection from the layer boundary to the starting point has the total angle of rotation equals to zero. Thus the reflection spectra of a chiral stratified structure are independent from the chirality parameter $\rho$ under normal wave incidence; a chiral structure and an achiral one with the same parameters are characterized by identical reflection spectra (one can compare curves of the reflection coefficient on Fig. 2 with the results given in [15]).

At oblique incidence of the exciting wave, the cross-polarized component appears in the reflected field too (Fig. 3). The cross-polarized components of the reflected and transmitted fields are equal to each other $|R^{eh}|=|R^{he}|$, $|\tau^{eh}|=|\tau^{he}|$, and at large angles of the incidence beginning from some cutoff angle, the magnitudes of the cross-polarized components of the reflected field and the co-polarized and cross-polarized components of the transmitted field decrease to zero ($|R^{ss'}|$, $|\tau^{ss}|$, $|\tau^{ss'}| \to 0$).

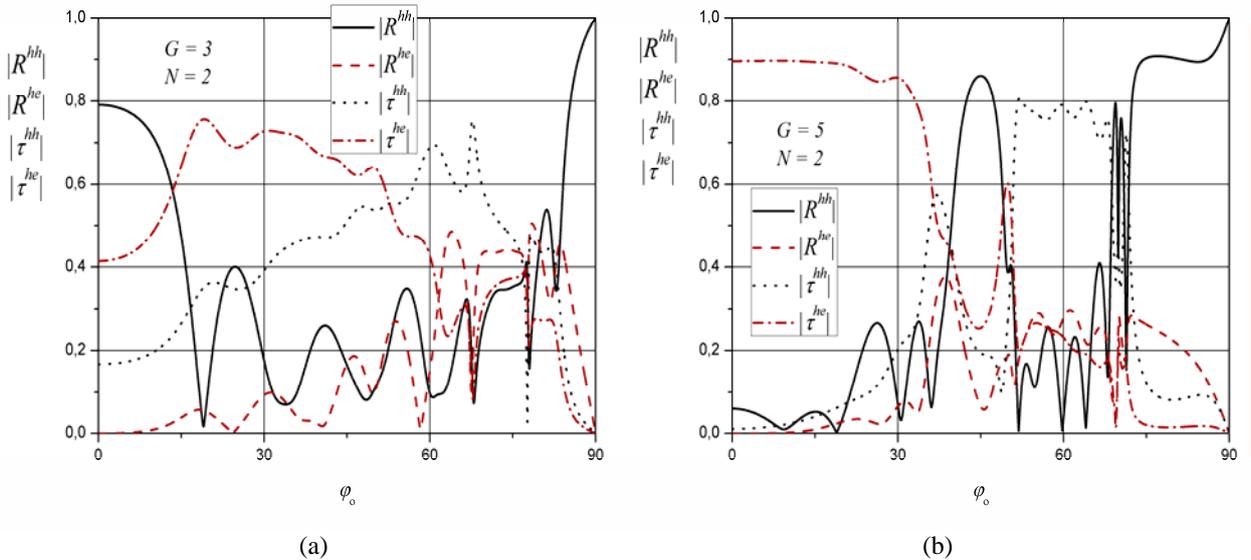

(a)          (b)

**Figure 3. (color online)** The amplitude of the reflection and transmission coefficients of a Cantor-like structure with chiral layers as a function of the angle of incidence $\varphi_0$ for different stages of fractal growth $(G, N)$.
$n_0 = n_1 = 1$, $n_2 = 1.5 + 0.013i$, $\rho = 0.1$, $\delta = 7$.

Further we focus our attention on the localized modes of the reflected field. As usually the case of normally incident plane waves is considered in previous papers [15-20]. As noted above, the features of the reflected field will depend crucially on the angle of wave incidence in case of a structure with chiral layers. Therefore the amplitude of the reflection coefficient of the fractal-like structure of convenient isotropic layers as a function of the thickness-to-wavelength ratio $\delta$ are given on Fig. 4 for a comparison. In order to clearly demonstrate all the peculiar properties of the spectra we choose the model materials with high refractive index contrast ($n_2/n_1 = 3$). A sample consists of 49 layers (24 and 25 layers with the refractive indexes $n_1$ and $n_2$, respectively) at the chosen fractal generation stage $(G=9, N=2)$. Thus the structure contains five multilayer inclusions separated by the gaps with thicknesses $L$ (Fig. 1).

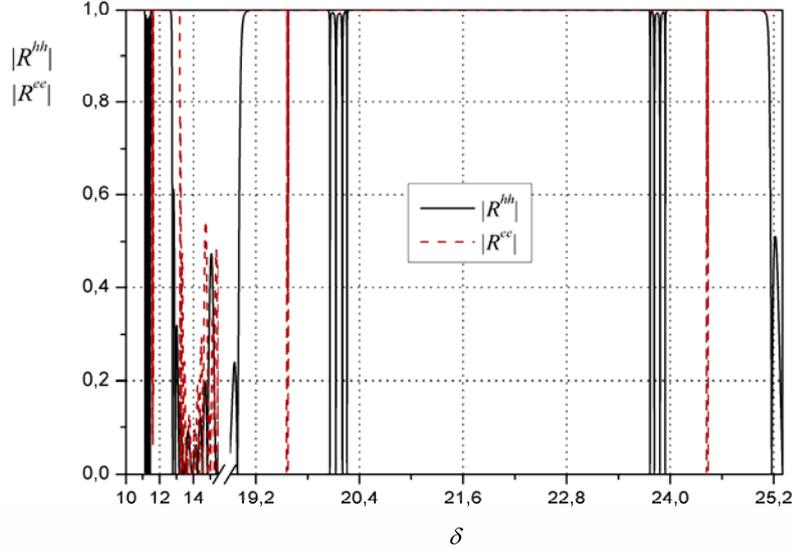

**Figure 4. (color online)** The amplitude of the reflection coefficient of a Cantor-like $(G=9, N=2)$ achiral structure ($|R^{he}|=|R^{eh}|=0$) as a function of the thickness-to-wavelength ratio $\delta$. $n_0 = n_1 = 1$, $n_2 = 3$, $\rho = 0.0$, $\varphi_0 = 45^0$.

Contrary to the case of normal wave incidence there is a difference of the reflection coefficient magnitudes of the orthogonally polarized waves, $|R^{hh}| \neq |R^{ee}|$. Additionally some symmetry distortion of the spectra appears within the period as result of the oblique wave incidence. Similar to the properties of defective periodic structures [26, 27] there are the sharp transmission resonances (peaks) inside the stop-bands. The position of these localized modes on the frequency scale is different for the orthogonally polarized waves. The peculiar property of the localized modes relatively to the fractal-like structure is their sequential splitting [18]. The last one appears as the interrelation between the generator $G$ and the number of peaks (single or multiplets) in the stop-bands. Thus the number of components in the multiplets equals to $(G-1)/2$, and the total number of peaks in one period equals to the number of layers, i.e., $G^N$. One can see from Fig. 4 that for chosen structure generation stage ($G=9$, $N=2$) there are four peaks in the leftmost and the rightmost stop-bands but in the central stop-band these peak multiplets are located symmetrically along the edges. The reason of such splitting is understood from the point of view of self-similarity of Cantor-like structures regarded in terms of coupled cavities [18]. The sample stack ($G=9$, $N=2$) contains five multilayer inclusions of ($G=9$, $N=1$) stacks. Each single peak in the spectrum of a ($G=9$, $N=2$) stack is a resonant mode produced by the material (yellow) layer ("cavity") with thickness $L$, splits into four modes because there are four such layers ("cavities") in a ($G=9$, $N=2$) stack.

The characteristic of the sequential splitting changes when a structure consists of chiral layers (Fig. 5). Additional peak multiplets appear in stopbands of both co-polarized and cross-polarized reflected fields. The position of these additional peaks is corresponding to the frequency of the localized modes of the orthogonally polarized wave in the achiral structure (Fig. 4). This effect is determined by the composition of the eigenmodes of the multilayer structure sections separated with the homogeneous gaps, and by additional eigenmodes that appear as a result of the wave polarization transformation [26]. The number of peaks in the additional multiplets is equal to the number of the peaks considered before for the achiral structure. The energy factor of the peaks and a distance between them within the multiplets depend on the chirality parameter $\rho$ periodically. Thus there is a possibility to change the localized mode configuration, i.e. to increase the energy factor of the additional peaks with reducing of that for the fundamental peaks [Figs. 5(a), 5(b)]. It is interesting in this context to note the suggestions of possible applications of the structure under study as optical switches and memories [14].

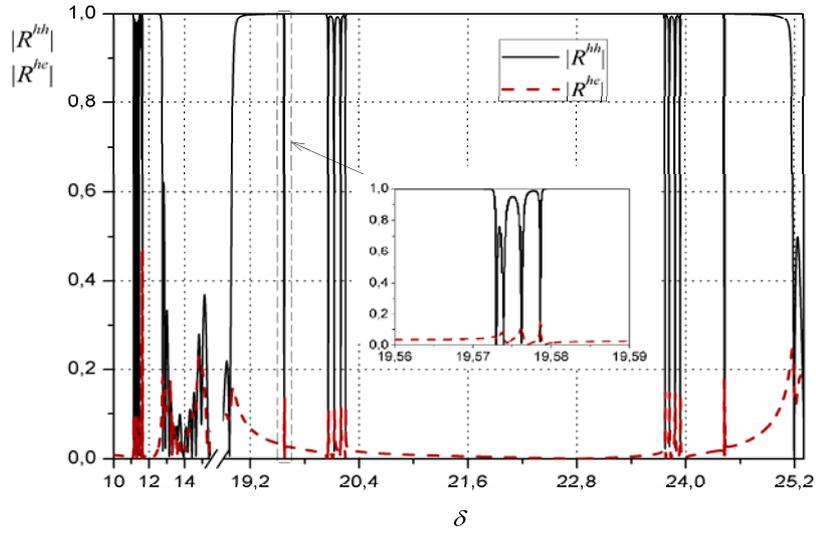

(a)

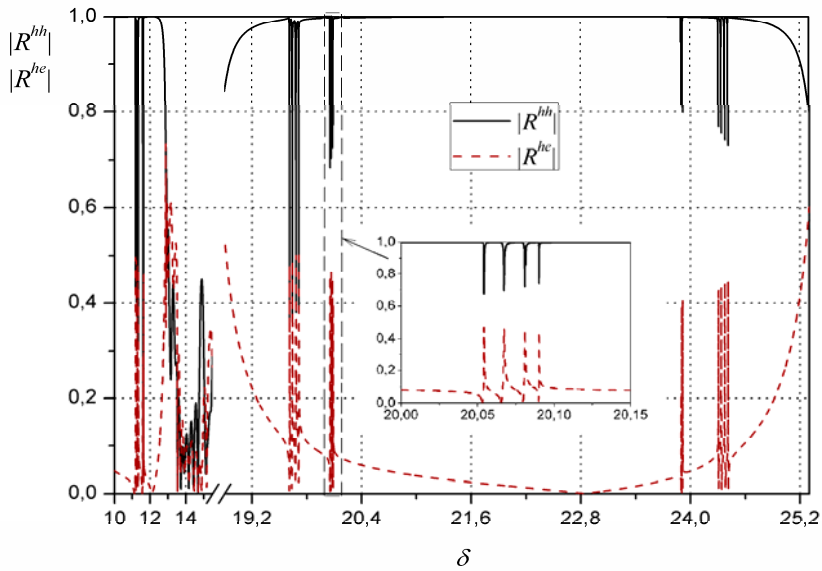

(b)

**Figure 5. (color online)** The amplitude of the reflection coefficient of a Cantor-like ($G = 9$, $N = 2$) structure with chiral layers as a function of the thickness-to-wavelength ratio $\delta$.
$n_0 = n_1 = 1$, $n_2 = 3$, $\varphi_0 = 45^0$. (a) $\rho = 0.05$, (b) $\rho = 0.3$.

## 5. Conclusion

The optical properties of the fractal-like stratified structure with chiral layers are studied. The method of solution is based on the $2\times 2$-block-representation transfer-matrix formulation.

Peculiarities of the wave localization, self-similarity, scalability and sequential splitting in the reflected and transmitted fields of the chiral multilayered structure are investigated in comparison with properties of achiral one. Especially the influence of the media chirality on properties of both the co-polarized and cross-polarized components of the reflected field is studied under oblique wave incidence.

The appearing of the additional localized modes (peak multiplets) in stopbands is revealed, and a correlation of their features with chirality parameter is established. The correlation between the geometrical and spectral properties of a Cantor-like multilayer structure is found out.

The revealed effects allow us to recommend the application of the studied structure in the design of optical switches and memories.